\documentclass{aa}

\usepackage[varg]{txfonts}
\usepackage{siunitx}
\DeclareSIUnit\um{\micro\meter}
\DeclareSIUnit\Msun{M_\odot}

\usepackage[x11names]{xcolor}
\usepackage{fancyvrb}
\usepackage{url}
\usepackage[tbtags,fleqn]{amsmath}
\usepackage{amsfonts}
\usepackage{graphicx}
\usepackage{natbib}
\usepackage{scalerel}

\newcommand{\Gravity}{\texttt{Gravity.jl}}

\DeclareMathOperator*{\LSE}{\scalerel*{\Lambda}{\sum}}


\begin{document}

\title{Gravity.jl: Fast and accurate gravitational lens modeling in Julia}
\subtitle{I. Point-like and linearized extended sources}
\author{Marco Lombardi}
\mail{marco.lombardi@unimi.it}
\institute{%
University of Milan, Department of Physics, via Celoria 16, I-20133
Milan, Italy} 
\date{Received ***date***; Accepted ***date***}

\abstract{%
We present \Gravity, a new proprietary software for the modeling of
gravitational lens systems. \Gravity\ is written in the Julia programming
language, and is designed to be fast, accurate, and flexible. It can be used
to model gravitational lens systems composed of multiple lensing planes, and
to perform Bayesian inference on the lens model parameters. In this paper we
present the theoretical and statistical ideas behind the code, and we describe
its main features. In this first paper of the series, we focus on the modeling
of point-like and small extended sources, for which we can linearize the lens
equation. We show a practical use of \Gravity\ on a galaxy-scale lens, and we
compare the results with those obtained with other codes. We also show how
\Gravity\ can be used to perform Bayesian inference on cosmological
parameters.} 
\maketitle

\section{Introduction}

The study of strong gravitational lenses, that is, massive objects such as
galaxies or clusters of galaxies capable of significantly distorting the light
coming from distant sources, represents both a unique opportunity and a
formidable challenge for astrophysics. Gravitational lensing analyses continue
to produce some of the most relevant scientific results in astrophysics. It
is, in fact, undoubtedly the most powerful tool we have for robustly studying
the total mass distribution of the galaxy or cluster of galaxies that act as a
lens, and therefore it allows us to place very strong constraints both on the
mass and on the possible presence of substructures through gravitational
imaging techniques. In this way, very relevant information can be obtained on
the properties of dark matter halos (see, e.g.,
\citealp{2010ARA&A..48...87T}). At the same time, the properties of the images
produced by a gravitational lens, especially if the lens and the source are at
redshift $z \gtrsim 0.3$, are influenced by the values of the cosmological
parameters, such as the mass density, the dark energy density, and equation of
state of the dark energy (see, e.g., \citealp{2022A&A...657A..83C}).
Therefore, systems with multiple images of sources at multiple redshifts, such
as galaxy clusters, are very well suited to be used for cosmological studies.
Moreover, in the last decade, it has been shown that in the case of variable
sources (such as quasars or supernovae), it is possible to carry out extremely
precise measurements of the value of the Hubble constant
\citep{2022A&ARv..30....8T, 2024A&A...684L..23G}. Finally, the magnification
produced by gravitational lenses allows us to observe and characterize objects
that would otherwise be too faint to be studied.

The study of gravitational lensing benefits enormously from the angular
resolution and optical stability offered by space telescopes. In particular,
new generation space telescopes such as Euclid and the \textit{James Webb
Space Telescope} (JWST) are formidable tools for studying gravitational
lensing. These telescopes, at the same time, are already producing a stream of
data whose interpretation poses nontrivial difficulties and challenges from a
computational point of view. In particular, just the modeling of multiple
point-image systems in lenses made up of galaxy clusters requires performing
Bayesian inference with many tens or hundreds of parameters. For example, the
lens modeling of a complex system such as the galaxy cluster MACS
J0416.1$-$2403, including 237 spectroscopically confirmed multiple images,
requires several weeks of computational time on a highly parallel workstation
\citep{2023A&A...674A..79B}. When the same operation is carried out on
extended sources in galaxy clusters, the complexity of the system is
beyond the capacity of the currently available strong lensing codes, unless
simplifications are used which severely limit the reliability of the results
obtained.

Added to these technical difficulties is a question that, from a statistical
point of view, is nontrivial. Several studies have shown that, even in ideal
cases (identical data, relatively simple systems) the results of strong
lensing analyses in galaxy clusters carried out by different researchers can
produce rather different results. The problem concerns the mass estimate to a
limited extent, provided the model is based on spectroscopic measurements of
several image families
\citep{2015ApJ...800...38G,2016A&A...587A..80C,2016ApJ...832...82J}, but instead
has a notable impact on other parameters such as the magnification of the
images \citep{2017MNRAS.465.1030P}. These uncertainties are typically
associated with larger-than-expected scatters between the predicted and observed
image positions, and are often attributed to the complexity of the lensing
system, to the presence of substructures, or to the presence of perturbers
along the line of sight \citep{2017MNRAS.465.1030P}. They also represent a
severe limitation to the use of strong gravitational lensing analyses for
cosmological studies.

Many of the discrepancies observed in the analyses of strong lensing systems
from different authors can be attributed to the different choices made in the
modeling of the lens system (in addition to further sources of uncertainty,
such as the associations of multiple images belonging to the same family).
Since many of the available codes are not able to evaluate the adequacy of
different models of a complex system such as a galaxy cluster in an effective
and statistically robust way, it is generally difficult to discriminate the
various models or take systematic uncertainties into account.

All these considerations suggest that there is a need for a new generation of
lensing codes, capable of performing fast and accurate lens modeling of
complex systems, and of performing Bayesian inference on the model parameters,
and expandable to allow the inclusion of new types of measurements. In this
paper, we present \Gravity, a new proprietary gravitational lensing modeling
software that aims to fulfill these requirements. \Gravity\ is written in
the Julia programming language, and is designed with speed and flexibility in
mind. In this first paper of the series, we focus on the modeling of
point-like and small extended sources, and we discuss the novel statistical
techniques used in the code.

For this software, we decided to use the Julia programming language
\citep{bezanson2017julia}. Julia is a high-level, high-performance programming
language for technical computing, with syntax that is familiar to users of
other technical computing environments. Although a complete discussion of the
features of Julia is beyond the scope of this paper, we believe that to
appreciate the advantages of \Gravity\ it is important to understand the main
features of the language. To the end user, Julia appears as an interpreted
language, much like Python, and similarly to Python, it allows one to use
multiple dispatch (that is, a function can accept multiple argument types).
However, Julia adopts just-in-time compilation techniques
\citep{10.1145/857076.857077}: the first time a function is called with some
specific parameter types, Julia compiles it to machine code, and then caches
the result. This means that Julia can achieve performance comparable to that
of C or Fortran while maintaining the flexibility and ease of use of an
interpreted language. Julia is also designed to be easy to use for parallel
and distributed computing, as well as for GPU-based computing. Most of these
advantages are associated with the use of the LLVM compiler infrastructure
\citep{LLVM}, which allows Julia to generate efficient machine code for a wide
range of architectures.

The paper is organized as follows. In Sect.~\ref{sec:theory} we present the
theoretical background, with a focus on the multiplane lensing equations and
on the Bayesian inference techniques used in \Gravity. In
Sect.~\ref{sec:measurements} we describe the various types of measurements
that can be used in the code for the various image parameters. In
Sects.~\ref{sec:image_plane_likelihood} and \ref{sec:source_plane_likelihood}
we write the forms of the likelihood function in the two possible schemes of
lensing analysis. The optimizations used for the computation of the likelihood
are described in Sect.~\ref{sec:image_plane_likelihood_optimizations}. In
Sect.~\ref{sec:technical_aspects} we describe the most important technical
aspects of the code and a few of the design choices. In
Sect.~\ref{sec:application} we present a practical use of \Gravity\ on a
galaxy-scale lens, and we compare the results with those obtained with other
codes. Finally, in Sect.~\ref{sec:conclusions} we draw our conclusions.

\section{Theoretical background}
\label{sec:theory}

\begin{figure*}
  \centering
  \includegraphics[width=0.8\textwidth]{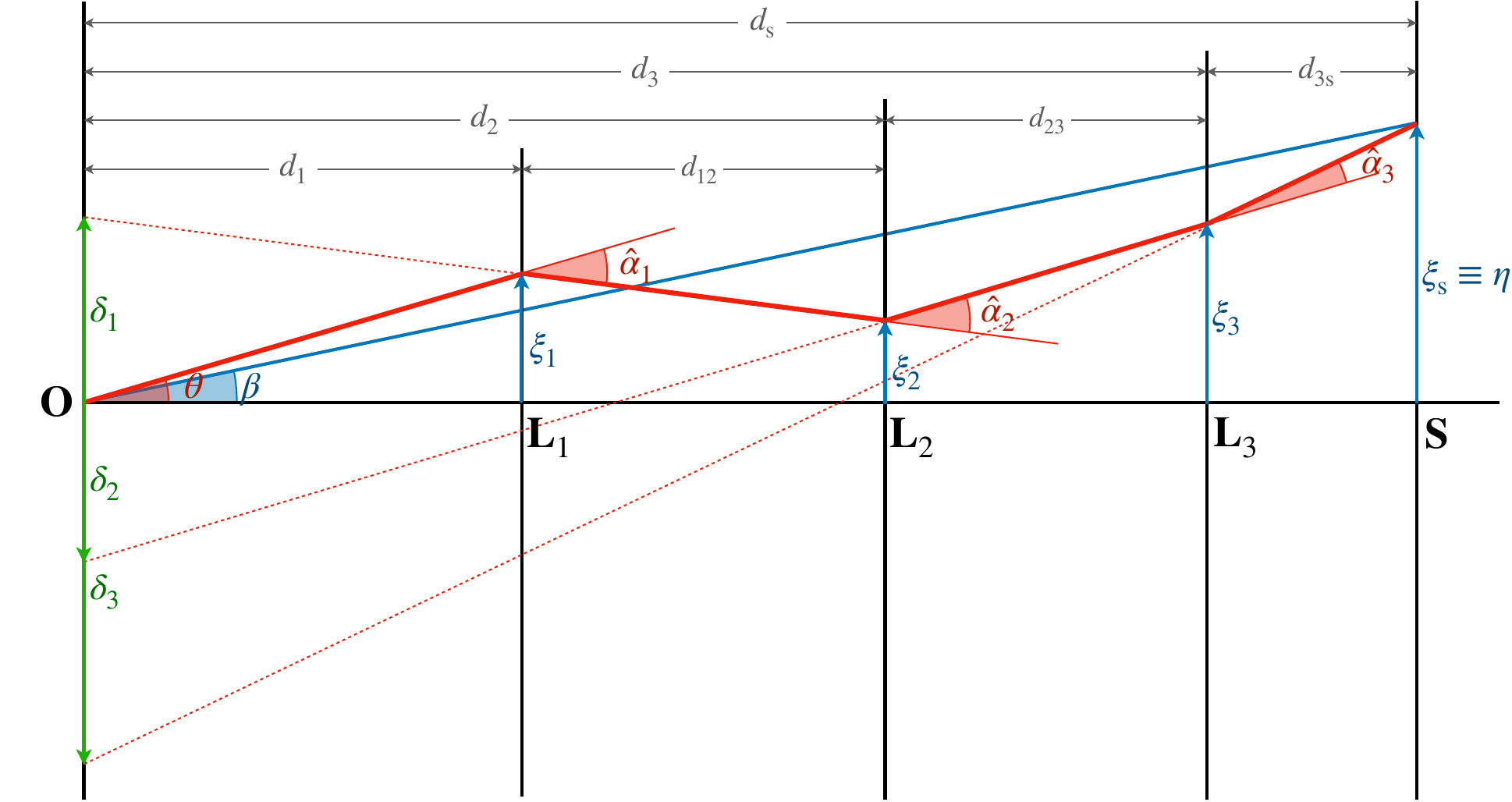}
  \caption{Schematic representation of a gravitational lens system. The observer
  is at the left, the source at the right. The lens is composed of multiple
  lensing planes, $\mathbf{L}_i$. A light ray (red) originating from the source
  plane $\mathbf{S}$ is deflected by the lensing planes and the associated
  image is formed at the observer's position along the direction $\theta$. The
  blue line in the direction $\beta$ is the position that the source would have
  in the absence of lensing. The deflection angles $\hat\alpha_i$ are computed at
  the angular positions $x_i$ of the images. All distances needed for the
  computation of the relation between $\theta$ and $\beta$ are marked at the top
  of the figure.}
  \label{fig:lens_system}
\end{figure*}

\subsection{Multiplane gravitational lensing}
\label{sec:multiplane}

\Gravity\ can perform lensing analyses of lens systems composed of multiple
lensing planes at different distances (redshifts). This capability is
implemented in terms of suitable distance coefficients and is based on the
multiplane lensing equations described below.

As discussed in the literature (see, e.g., \citealp{1992A&A...265....1S} and
\citealp{2009A&A...499...31H}), multiplane lensing equations can be written
iteratively, essentially following backward the light path from the
observer to the source. A light ray, during his path, is subject to many
deflections at its intersection with the various lensing planes. These
deflections are often described in the literature in terms of the
reduced deflection angle $\alpha_i = d_{is} \hat \alpha_i / d_s$,
where $d_{is}$ (respectively $d_s$) is the distance between the $i$-th lens
plane and the source (respectively, the observer and the source), and $\hat
\alpha_i$ is the true deflection angle at the $i$-th lensing plane. The last
quantity is actually the physical deflection and depends solely on the mass
distribution at the lens plane, with no dependence on any distance.

Since \Gravity\ needs to deal with many lens planes and many sources at
different distances, it is much more efficient to write multiplane lens
equations in terms of $\hat \alpha$ (rather than $\alpha$, as is done
usually). To this purpose, let us call $x_i$ the angular position of the
intersection of the light path with the $i$-th lensing plane, as seen from the
observer, and $\xi_i$ the corresponding position in the lensing plane in
comoving coordinates (see Fig.~\ref{fig:lens_system}). We have then $x_i =
\xi_i / d_i$, and $x_1 \equiv \theta$ corresponds to the angular position of
the image as seen from the observer. We call $\delta_i$ the vector denoting
the intersection of the extension of the $i$ light path segment (that is, the
path that connects the plane $\mathbf{L}_{i-1}$ with the plane $\mathbf{L}_i$)
with the plane containing the observer $\mathbf{L}_0 \equiv \mathbf{O}$. By
definition, we have $\delta_1 = 0$. We find then
\begin{align}
  \delta_{i+1} &{}= \delta_i + d_i \hat \alpha_i(x_i) \; , &
  x_{i+1} &{}= x_i - \rho_i \delta_{i+1} \; .
\end{align}
Here $\rho_i = d_{i,i+1} / d_i$ is the ratio of the distance between the
$i$-th plane and the next $(i+1)$-th plane and the distance of the $i$-th
plane to the observer. We again note that all distances here and below are taken
to be comoving transverse cosmological distances, rather than
angular-diameter distances as usually done: since distance ratios are
involved, this distinction is important only for the computation of the
time delay below.

The advantage of using this approach is twofold. On the one hand, the
computation of the deflection angles $\hat \alpha_i$ can be done in a way that
is independent of the distances, and therefore one can focus on the physical
properties of the specific lens (instead of a combination of physical
parameters and cosmological distances). Additionally, in case the distance
of a lens or source plane is taken as a free parameter, the code needs to
recompute just two (if the plane is the first or the last) or three distances
(in the other cases), instead of a plethora of distances that would be needed
if the deflection angles were computed in terms of $\alpha$. Finally, when the
cosmological model is not known, the computation scheme used in \Gravity\
reduces the number of cosmological distances to compute at each change of the
cosmological parameters: in the case of Fig.~\ref{fig:lens_system}, for
example, one would need to compute 7 distances, instead of 10.

The differential of previous equations gives the Jacobian matrix of the
lensing equation (we note that in the case of multiplane lensing the Jacobian
matrix is often not symmetric). We define
\begin{align}
  \Delta_1 &{}= \begin{pmatrix} 0 & 0 \\ 0 & 0 \end{pmatrix} \; , &
  A_1 &{}= \begin{pmatrix} 1 & 0 \\ 0 & 1 \end{pmatrix}
\end{align}
as the null and identity matrices. We then have
\begin{align}
  \Delta_{i+1} &{}= \Delta_i + d_i \frac{\partial \hat \alpha_i(x_i)}{\partial x_i} A_i \; , &
  A_{i+1} &{}= A_i - \rho_i \Delta_{i+1} \; .
\end{align}
Finally, the time delay expression can be computed as
\begin{equation}
  T = \frac{1}{c} \sum_i \frac{1}{2} \rho_i \delta_{i+1}^2 - d_i \psi_i(x_i) \; ,
\end{equation}
where $\psi_i(x_i)$ is the Fermat's potential of the $i$-th lens plane
computed at the angular position $x_i$.

\subsection{Bayesian inference}
\label{sec:bayes}

Bayes' theorem (see, e.g., \citealp{MacKay}) is at the base of the inference
approach adopted in \Gravity. In its standard form, it is used to compute the
probability of a set of parameters $x$ of a model $M$ given the
(observational) data $D$. It can be written as
\begin{equation}
  \label{eq:bayes}
  P(x \mid D, M) = \frac{P(D \mid x, M) P(x \mid M)}{P(D \mid M)} \; \cdot
\end{equation}
The quantity $P(x \mid D, M)$ is posterior distribution of $x$ given $D$, and
is generally the final result we are interested in. It depends on $P(D \mid x,
M)$, that is, the likelihood of the data given the parameters, and on $P(x \mid
M)$, the so-called the prior distribution of the parameters. The likelihood
generally encapsulates both the physical model (in our case, a multiplane
lensing system) and the statistical properties of the measurements. The prior,
instead, is a description of our beliefs on the parameters $x$, and it plays a
critical role in Bayesian statistics. It should be fixed before looking at the
data and should for no reason be influenced by the data. The prior is often
the most controversial aspect of Bayesian statistics, and it is often the
source of criticism of the Bayesian approach. In fact, the need to specify a
(proper) prior is a guarantee of reproducibility of statistical inferences, as
people using the same model and assuming the same prior distribution reach
the same conclusions. It is important also to distinguish between the choice
of a prior for the unknown model parameters, and some initial guess that one
might make for the same parameters. We see below how \Gravity\ helps to
distinguish these two concepts. 

The quantity $P(D \mid M)$, often called the evidence, can be expressed as a
marginalized likelihood:
\begin{equation}
  P(D \mid M) = \int P(D \mid x', M) P(x' \mid M) \, \mathrm{d} x' \; ,
\end{equation}
where the integral is carried out over the relevant domain of the parameters
$x$. The evidence is a normalization factor that ensures that the posterior is
a proper probability distribution. Its computation requires the use of a
proper, normalized prior:
\begin{equation}
  \label{eq:prior_norm}
  \int P(x' \mid M) \, \mathrm{d} x' = 1 \; .
\end{equation}
It is usually very difficult to compute the evidence analytically, and it is
often ignored in practice. It is, however, the single most important quantity,
and it has a key role in the assessment of the goodness of a model in the
Bayesian framework. Specifically, suppose that one wishes to evaluate $P(M
\mid D)$, that is, the probability that the model $M$ is correct, given the
data $D$. Using Bayes' theorem, we can express this probability in terms of
the evidence $P(D \mid M)$:
\begin{equation}
  P(M \mid D) = \frac{P(D \mid M) P(M)}{P(D)} \; \cdot
\end{equation}
In this expression $P(M)$ is a prior over $M$, that is, our belief that $M$ is
the correct model before looking at the data. The normalizing term $P(D)$ is
often unknown: to compute it, one would need to marginalize over all possible
models, which is generally impossible. This prevents us from estimating $P(M
\mid D)$, but not the ratio $P(M \mid D) / P(\tilde{M} \mid D)$, where
$\tilde{M}$ is an alternate model:
\begin{equation}
  \frac{P(M \mid D)}{P(\tilde{M} \mid D)} = \frac{P(D \mid M)}{P(D \mid \tilde{M})}
  \frac{P(M)}{P(\tilde{M})} \; \cdot
\end{equation}

For this reason, in \Gravity\ we provide various tools to evaluate the
evidence.

\subsubsection{Prior and initial guesses}
\label{sec:prior_initial_guesses}

In \Gravity, we distinguish among several probability distributions:

\begin{description}
  \item[\textit{Prior.}] It is a description of our beliefs on the model
  parameters, and it is used to compute the posterior distribution through
  Bayes' theorem. This distribution should be fixed before looking at the data,
  and should not be influenced by the data. The prior should be normalized to
  unity as described in Eq.~\eqref{eq:prior_norm}.
  \item[\textit{Reference distribution.}] Often, the algorithms used to
  investigate the posterior (generally based on a sampling of the
  posterior through Markov chain Monte Carlo techniques), require one or more
  initial guesses of the parameters. For example, algorithms such as
  Metropolis-Hasting or ensemble samplers (such as the affine-invariant sampler
  of \citealp{2010CAMCS...5...65G}) require a set of initial guesses to start
  the sampling. These initial guesses can, but do not need to be, generated
  using the prior distribution: they are just a set of points in the parameter
  space where the sampler starts its exploration, and as such one can tailor
  them using the data (while, as stressed above, the prior should never
  make use of the data). In \Gravity, we distinguish strictly between the prior,
  and the distribution used to generalize the initial guesses, which we call the
  reference distribution.
  \item[\textit{Posterior.}] The posterior distribution of the model parameters. This
  distribution is the result of the application of Bayes' theorem to the prior
  and the likelihood. It is our belief on the model parameters after updated
  according to the data.
\end{description}

A typical use of these distributions is the following. In a first step, one
might perform an initial, simplified analysis of a lens system, for example
using a source-plane analysis (see below
Sect.~\ref{sec:source_plane_likelihood}). For this analysis, one would set a
prior distribution over the lens parameters, and use it to compute the
posterior distribution. This posterior distribution can then be used to build
the reference distribution, which is used to generate the initial guesses for
a more complex image-plane analysis. Since the image-plane analysis makes use
of additional source parameters, the reference distribution needs to be
complemented with educated guesses of the source parameter distribution.
Finally, this reference distribution, together with an image-plane prior, can
be used to start the sampling of the final image-plane posterior distribution
of the system. Although all these steps are common practice in strong
gravitational lensing, they are implemented in \Gravity\ in a way that is as
transparent as possible to the user and that strictly satisfies the
requirements of Bayesian statistics.

\subsubsection{A note on marginalization}

When performing a Bayesian analysis of a complex system, we are often
uninterested in a set of nuisance parameters that are needed in the modeling
of the problem. A relevant example in our context is the case of source
parameters in the modeling of a gravitational lens system.

In general, we can consider the likelihood associated with a gravitational
lens system as formed by (at least) two sets of parameters: source parameters
$S$ (in the simpler case, the positions of all sources; in more complicated
cases also their luminosities and shapes) and lens parameters $L$. In more
complex case we might have more parameters, for example, associated with the
cosmological model, but for simplicity in this discussion we ignore these
cases.

Let us call $D$ the data obtained (for example $D = \{ \hat\theta_n \}$ if
we limit our analysis to the image positions). We write the likelihood, that is,
the conditional probability of the data given the parameters, as $P(D \mid S,
L)$. This quantity is then used in Bayes' theorem so that we have
\begin{align}
  P(L \mid D) & {} = \int P(S, L \mid D) \, \mathrm{d}S \notag\\ 
  & {} = 
  \frac{P(L) \int \mathrm{d}S \, P(S) P(D \mid S, L)}
  {\int \mathrm{d}L' \, P(L') \int \mathrm{d}S' \, P(D \mid S', L') P(S')}
  \; \cdot
\end{align}
As a result, if we write the conditional distribution $P(D \mid L)$ as
a marginalization over $S$ of the full likelihood,
\begin{equation}
  P(D \mid L) = \int P(D \mid S, L) P(S) \, \mathrm{d}S \; ,
\end{equation}
we see that we can recover the usual form of Bayes' theorem
\begin{equation}
  P(L \mid D) = \frac{P(D \mid L) P(L)}{\int P(D \mid L') P(L') \mathrm{d}L'} \; \cdot
\end{equation}
In a sense, this marginalization corresponds to the computation of a partial
evidence over the source position. The same result can be obtained by
considering the definition of the conditional probability:
\begin{align}
  P(D \mid L) & {} = \frac{P(D, L)}{P(L)} = 
  \frac{\int P(D, L \mid S) P(S) \, \mathrm{d}S}{P(L)} \notag\\
  & {} =
  \int \frac{P(D, L \mid S)}{P(L)} P(S) \, \mathrm{d}S = 
  \int P(D \mid S, L) P(S) \, \mathrm{d}S \; .
\end{align}
Therefore, it is sensible to compute this marginalized conditional
distribution. This can be done by adopting the technique described below in
the paper.

\subsubsection{Conjugate priors}

It is sensible to assume that the measurements of the parameters
characterizing our point-images, such as their position, their luminosity, or
their shape, follow simple probability distributions. For example, position
measurements can be taken to be distributed as bi-variate Gaussian. In these
situations, with a suitable choice of the prior (using the so-called conjugate
prior), we can make sure that the posterior belongs to the same family of the
prior. This greatly simplifies the calculations and allows us to compute
analytically the evidence required, as explained below.

\section{Measurements}
\label{sec:measurements}

\Gravity\ is designed to work with a variety of measurements. In this section
we describe the various types of measurements that can be used in the code for
the various image parameters. We stress that, since in this paper we
essentially focus on the modeling of point-like sources (with possible
extensions such as luminosity and time delay measurements), we do not need to
consider here the effects of the point-spread function (PSF) of the telescope.
This effect is instead crucial for the study of extended sources, and will be
discussed in a future paper. 

\subsection{Point-like sources}

Image measurements are at the core of any lensing inversion. In \Gravity\ we
assume everywhere that point-like image measurements have normal (Gaussian)
errors: more precisely, we write the probability of observing an image at the
position $\hat\theta$ given that its true position is $\theta$ as a
bivariate normal distribution with a given known precision matrix
$\mathbf\Theta$ (this matrix is the inverse of the covariance matrix, a $2
\times 2$ symmetric matrix representing the measurement errors):
\begin{equation}
  P(\hat\theta \mid \theta) = \sqrt{\biggl| \frac{\mathbf\Theta}{2\pi} \biggr|}
  \exp\left[ -\frac{1}{2} (\hat\theta - \theta)^T \mathbf\Theta (\hat\theta - \theta) 
  \right] \; .
\end{equation}
We also assume that all image measurements are independent: as a result, when
computing the joined probability to observe a set of images $\{ \hat\theta_i
\}$ given the corresponding predictions $\{ \theta_i \}$, we can just write
\begin{equation}
  P(\{ \hat\theta_i \} \mid \{ \theta_i \}) = \prod_{i=1}^I
  \sqrt{\biggl| \frac{\mathbf\Theta_i}{2\pi} \biggr|}
  \exp\left[ -\frac{1}{2} (\hat\theta_i - \theta_i)^T 
  \mathbf\Theta_i (\hat\theta_i - \theta_i) \right] \; .
\end{equation}
This is effectively the likelihood generally used in the code.

\subsection{Luminous sources}

In case where we also measure the images' magnitudes $\{\hat m_i\}$ and their
associated inverse variances $\{\lambda_i\}$ we can easily add the related
constraints to the previous equations. The likelihood associated with the
magnitudes is simply
\begin{equation}
  P(\{ \hat m_i \} \mid \{ m_i \}) = \prod_{i=1}^I \sqrt{\frac{\lambda_i}{2\pi}}
  \exp\left[ -\frac{1}{2} \lambda_i (\hat m_i - m_i) \right] \; .
\end{equation}

The observed magnitudes are related to the original (unlensed) one $M$
through the relation
\begin{equation}
  m_i = M - 2.5 \log_{10} \bigl\lvert A_i^{-1} \bigr\rvert \equiv M - \mathit{LM}_i \; ,
\end{equation}
where we have called $\mathit{LM}_i \equiv 2.5 \log_{10} \bigl\lvert A_i^{-1}
\bigr\rvert$ the lensing modulus, a quantity that indicates the change in
magnitude associated with the lensing magnification. With this definition, we
can write the likelihood as
\begin{equation}
  P \bigl(\{ \hat m_i \} \mid M \bigr) = \prod_{i=1}^I \sqrt{\frac{\lambda_i}{2\pi}}
  \exp\left[ -\frac{1}{2} \lambda_i \bigl(\hat m_i + \mathit{LM}_i - M \bigr)^2
  \right] \; .
\end{equation}

\subsection{Elliptical sources}

As a third source type, we consider sources with an elliptical profile. These
sources can be characterized using the quadrupole moment of their light
distribution: for a source with semi-axes $a$ and $b$ and position angle
$\varphi$ counted clockwise from the top (north to west in astronomical
sense), the quadrupole moment is
\begin{equation}
  Q = \begin{pmatrix}
    a^2 \sin^2 \varphi + b^2 \cos^2 \varphi & (a^2 - b^2) \sin \varphi \cos \varphi \\
    (a^2 - b^2) \sin \varphi \cos \varphi & a^2 \cos^2 \varphi + b^2 \sin^2 \varphi
  \end{pmatrix} \; . 
\end{equation}
We note how the quadrupole moment is a symmetric and positive-definite matrix. We
take quadrupole moment measurements to be unaffected by the PSF (or, more
realistic, we assume that the effects of the PSF have been removed). We also assume
that they are distributed according to a Wishart
distribution (see, e.g., \citealp{Livan2018-za}), with probability density
function given by
\begin{equation}
  P(\hat Q \mid \mathbf{W}, \nu) = \frac{|\mathbf{W}|^{\nu/2}}{2^{\nu p/2} \Gamma_p(\nu/2)}
  |\hat Q|^{(\nu-p-1)/2} \exp\bigl[ - \mathrm{tr}(\mathbf{W} \hat Q) / 2\bigr] \; ,
\end{equation}
where $p = 2$ is the dimensionality of the space, $\nu > p + 1$ is a real
number, and $\mathbf{W}$ is a symmetric, positive definite matrix. The
equation above also uses $\Gamma_p$, the multivariate Gamma function. For
the specific case $p = 2$ we have
\begin{equation}
  \Gamma_2 (\nu/2) = \pi 2^{2-\nu} \Gamma(\nu-1) \; .
\end{equation}
Therefore, for $p = 2$ the probability distribution function simplifies into
\begin{equation}
  P(\hat Q \mid \mathbf{W}, \nu) = \frac{|\mathbf{W}|^{\nu/2}}{4 \pi \Gamma(\nu-1)}
  |\hat Q|^{(\nu-3)/2} \exp\bigl[ - \mathrm{tr}(\mathbf{W} \hat Q) / 2 \bigr] \; .
\end{equation}
The mode of this distribution is $(\nu - p - 1) \mathbf{W}^{-1} = (\nu-3)
\mathbf{W}^{-1}$, while the average is $\nu \mathbf{W}^{-1}$. As usual, we
assume that all measurements are independent, so that the joint distribution
for various elliptical measurements can be obtained as a product of terms such
as the one written above. We note that here we use the canonical parameter
$\mathbf{W}$ instead of the more usual choice $\mathbf{V} \equiv
\mathbf{W}^{-1}$.

A quadrupole measurement is generally given in terms of a measured shape,
often given in terms of the semi-axes ($a$ and $b$) and of a position angle
($\theta$), together with their associated errors ($\sigma_a$, $\sigma_b$,
$\sigma_\theta$). A quadrupole moment $\hat Q$ can be easily built using the
set $\{ a, b, \theta \}$ from the equations above. We see below how we
model the intrinsic quadrupole of a source and its relation to the
$\mathbf{W}$ matrix. Regarding $\nu$, it is easy to show that, to first order,
one has
\begin{equation}
  \frac{\sigma_a}{a} = \frac{\sigma_b}{b} = \frac{2}{\sqrt{\nu}} \; ,
\end{equation}
where $\sigma_a$ and $\sigma_b$ are the measurement uncertainties on the
semi-axes $a$ and $b$.

\subsection{Time delays}

In some cases, we have at our disposal time delays for the multiple images of
a point source, and we want to take advantage of this information. This is
particularly useful for cosmological measurements, in particular the ones
related to the Hubble's constant.

The situation here, from a statistical point of view, is virtually identical
to the case of magnitude measurements. Specifically, suppose we measure the
images' time delays $\{ \hat t_i\}$ and their associated inverse variances
$\tau_i$. We note that, typically, time delays are given as differences in
arrival time concerning a given source. In \Gravity\ we assume that all images
of a given source are given a time delay: this, in practice, means that one
usually sets the measured time delay of the reference image to zero with a
very small error (i.e., a very large precision $\tau_i$).

The computation of time delays requires the knowledge of Fermat's potential,
which for many lens models is a relatively expensive operation. From a
statistical point of view, however, all equations for time delay measurements
reflect the ones for magnitude measurements. Throughout the code we assume
that the time delay values $\hat t_i$ are given with respect to a reference
event: for example, we could set the time delay of a reference image to
zero, and measure the time delays of the other images with respect to this
reference. The likelihood for the time delays can then be written as
\begin{equation}
  P(\{ \hat t_i \} \mid \{ t_i \}) = \prod_{i=1}^I \sqrt{\frac{\tau_i}{2\pi}}
  \exp\left[ -\frac{1}{2} \tau_i (\hat t_i - t_i) \right] \; ,
\end{equation}
with the expected time delays $\{ t_i \}$ written as
\begin{equation}
  t_i = T + T_i \; .
\end{equation}
Here we have called $T_i$ the time delay function for the $i$-th image, and
$T$ the reference time for the source.

\subsection{Log-likelihoods}

Generally, we are dealing with the log-likelihood, which therefore can be
written as
\begin{equation}
  \log P(\{ \hat\theta_i \} \mid \{ \theta_i \}) = \sum_{i=1}^I \biggl[ \frac{1}{2} 
  \log \biggl| \frac{\mathbf\Theta_i}{2\pi} \biggr| - 
  \frac{1}{2} (\hat\theta_i - \theta_i)^T \mathbf\Theta_i (\hat\theta_i - \theta_i)
  \biggr] \; ,
\end{equation}
for the positions,
\begin{equation}
  \log P \bigl(\{ \hat m_i \} \mid M \bigr) = \sum_{i=1}^I \biggl[ \frac{1}{2} 
  \log \frac{\lambda_i}{2\pi}
  - \frac{1}{2} \lambda_i \bigl(\hat m_i + \mathit{LM}_i - M \bigr)^2
  \biggr] \; ,
\end{equation}
for the magnitudes (and similarly for the time delays), and
\begin{align}
  & \log P \bigl(\{ \hat Q_i \} \mid \{ \mathbf{W}_i \}, \{ \nu_i \} \bigr) =
  \sum_{i=1}^I \biggl[ \frac{\nu_i}{2} \log |\mathbf{W}_i| 
  - \log \bigl[ 4 \pi \Gamma(\nu_i - 1) \bigr] + {} \notag\\
  & \qquad \frac{\nu_i - 3}{2} \log |\hat Q_i| - 
  \frac{1}{2} \mathrm{tr} (\mathbf{W}_i \hat Q_i)
  \biggr] \; ,
\end{align}
for the quadrupole moments.

In a maximum-likelihood approach, in many cases one can safely ignore the
normalizing constants in the expressions above, corresponding in all cases to
the first two terms inside the summations, as their values typically depend
only on the measurement errors associated with each observed image, and not on
the predicted images. In a Bayesian approach, instead, the normalizing
constants can be relevant, especially when a marginalization over the source
parameters is performed, or when one wishes to compare different lensing
models.

When dealing with these expressions, we need to consider various issues
\begin{enumerate}
  \item How are the predicted images $\{ \theta_i \}$ computed?
  \item How do we associate an observed image to the corresponding predicted
  one?
  \item What do we do if the number of predicted images differs from the number
  of observed ones?
\end{enumerate}
We consider all these issues below.

\section{Image-plane likelihood}
\label{sec:image_plane_likelihood}

The most straightforward way to compute the likelihood of a lens model is to
include explicitly the source parameters (positions, magnitudes, quadrupole
moments...) in the model, and to perform an inversion of the lens mapping and
associate the counter images of a given source. All these operations, as noted
above, are nontrivial and time consuming. For this reason, usually, an
image-plane optimization is carried out only after an approximate lens model
has been found using a source-plane optimization (see below
Sect.~\ref{sec:source_plane_likelihood}).

In principle, once the counter-images of a given source have been found, the
computation of the image-plane likelihood is a mere application of the
equations written in the previous section. In practice, things are a little
more complicated because (1) the inversion of the lens equation might produce
too many or too few predicted images, and (2) the predicted images might not
be associated with the observed ones in a one-to-one way. We consider
these issues below. In the following we reserve the subscript $i \in \{1,
\dots, I\}$  to indicate the observed images and the subscript $p \in \{1,
\dots, P\}$ for the predicted images from a given source.

\subsection{Inversion of the lens equation}

\begin{figure}
  \centering
  \includegraphics[width=\linewidth]{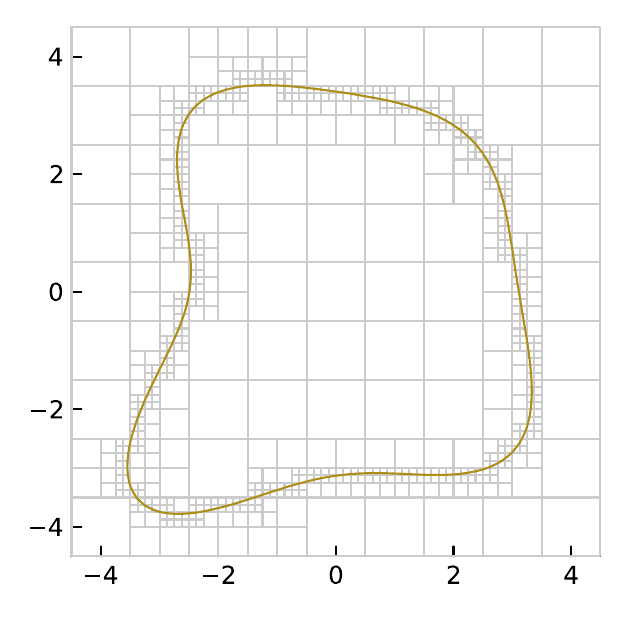}
  \caption{Example of grid with an order=3 refinement near a critical curve.}
  \label{fig:grid_refinement}
\end{figure}

The inversion of the lens equation, that is, the discovery of the image
positions associated with a given source position, plays a central role in the
image-plane likelihood. This task is generally carried out using a nonlinear
equation solver. Many nonlinear equation solvers are known, but in \Gravity\
we have decided to use the simplest one, Newton–Raphson's method.

Newton's method, like any iterative method, requires a starting guess of the
solution. When we have already a good approximation of the lens model, we can
generally safely use the positions of the observed images as starting points
of the method. The convergence of the lens inversion method is in these cases
generally very fast, unless the Jacobian of the lens equation is not
well-behaved (this can happen for sources on top of critical curves with
formally infinite magnification, or for lens models with singularities).

In case we do not have a good lens model, or in case we wish to find new
solutions of the lens equations (for example, to find possible
counter-images), we can use a suitable grid and map all grid cells into the
source plane using the lens equation. We can then identify the cells that,
when mapped in the source plane, enclose the source position, and use
their centers (or better, the inverse of the source position using a linear
approximation of the lens equation) as guesses for the solution. If the grid
is suitable built, this technique generally ensures that we find all
possible solutions of the lens equation, at the expense of a much higher
computational complexity. This task, which can be easily parallelized, is
optionally associated with a hierarchical binary refinement of the grid cell
near critical lines, observed images, or centers of the lens profiles, to
ensure as much as possible that all possible solutions are found.
Figure~\ref{fig:grid_refinement} shows an example of such a grid refinement.

Our grid algorithm has been designed to be as general and as robust as
possible. Critical lines are refined using a simple criterion based on the
sign of the determinant of the Jacobian at the vertices of each cell
(following a prescription also adopted by \citealp{2007NJPh....9..447J}): When
the sign changes between the two vertices of a cell side, the cell is refined.
The algorithm proceeds then recursively, refining the cells that are found to
be on the critical lines. We note that during this refinement procedure, it might
happen that an unrefined cell is found to be on a critical line: in this case,
the cell is refined as well. This latter point is important, as it guarantees
that critical lines are always resolved with the desired accuracy and also
avoid artifacts when drawing critical lines (see, e.g., Fig.~A2 of
\citealp{2007NJPh....9..447J} or Fig.~1.1 of the manual of Glafic~2,
\citealp{2010PASJ...62.1017O, 2021PASP..133g4504O}).

\subsection{Image positions}

In the simplest case, the log-likelihood for a given family of predicted images
includes only the image positions. This case, considered here, shows already
several interesting features.

\subsubsection{Direct association}

If, somehow, we have sorted out the predicted images so that each of them is
associated with the corresponding observed image (which necessarily implies $I
= P$), we can just compute the log-likelihood as
\begin{equation}
  \label{eq:logl_pos_ip}
  2 \log P(\{ \hat\theta_i \} \mid \{ \theta_i \}) =
  \sum_{i=1}^I \log{\biggl| \frac{\mathbf{\Theta}_i}{2\pi} \biggr|} - 
  (\hat\theta_i - \theta_i)^T \mathbf{\Theta}_i (\hat\theta_i - \theta_i)
  \; .
\end{equation}
The log-likelihood expression above can be easily differentiated with
respect to the source position $\beta$: taking into account again the
theorem of the inverse function, we find
\begin{equation}
  \frac{\partial \log P(\{ \hat\theta_i \} \mid \{ \theta_i \})}{\partial \beta}
  = \sum_{i=1}^I A_i^{-T} \mathbf{\Theta}_i (\hat\theta_i - \theta_i) \; .
\end{equation}
This quantity can be useful when performing a likelihood maximization over the
source position, or when using Monte Carlo techniques that require the
derivatives of the posterior (such as Hamiltonian Monte Carlo).

\subsubsection{Best match}

In many cases, we cannot associate directly one observed image with one
predicted image. This can happen, for example, when the lens model is not
accurate enough, or when the source is very close to a critical curve. In
these cases, we might decide to find the best association between the observed
and predicted images, and use this association to compute the likelihood.

Let us call $\sigma$ the permutation that associates each observed image to
one predicted image. We can then write the log-likelihood as
\begin{equation}
  2 \log P(\{ \hat\theta_i \} \mid \{ \theta_p \}, \sigma) =
  \sum_{i=1}^I \log{\biggl| \frac{\mathbf{\Theta}_i}{2\pi} \biggr|} - 
  (\hat\theta_i - \theta_{\sigma(i)})^T \mathbf{\Theta}_i (\hat\theta_i - \theta_{\sigma(i)})
  \; ,
\end{equation}
We use different subscripts for the images and predictions because we have, in
general, different numbers of them. The procedure finds the permutation
$\sigma$ that maximizes the expression above and returns the maximum value
found, together with the Jacobian with respect to the source position.

The exact meaning of ``permutation'' here depends on a specific choice:
\begin{itemize}
\item We might use for $\sigma$ permutations with repetitions: this way, each
observed image can be matched to each predicted one with no restriction, so
that two observed images could be in principle associated with the same
predicted one. This way we have $P^I$ different permutations. In practice,
finding the best permutation is very simple: we just compute all possible sum
terms above and take the best ones for each source. Thus, essentially, for
each observed image we take the closest predicted image in terms of
Mahalanobis distance \citep{Mahalanobis}.
\item Alternatively, we might use permutations without repetitions: in this
case one predicted image can be associated with one observed image at most,
resulting in ${}^P \mathrm{P}_I = P! / (P - I)!$ possible permutations. In
this case, finding the best permutation is more complicated, as we need to
check that we do not have repetitions in the best solution. The strategy we
adopt to find the best $\sigma$ is to try the same technique as in the case of
permutations with repetition, and check if we have repetitions in the best
solution: if we do not, we take this solution; if we do, we loop over the
(possibly large) number of permutations and take the best one. This choice
requires that the predicted images are at least as numerous as the observed
ones, so that $P \ge I$. This is potentially problematic, as it might
prevent a convergence if the lensing model is far from a realistic
configuration.
\end{itemize}

\subsubsection{Marginalization over all matches}

Looking again at the previous subsection, we can easily recognize that the
permutation $\sigma$ is a nuisance parameter: we are not really interested in
knowing its value, as we merely use it to compute the likelihood. In a more
correct Bayesian approach, we should therefore marginalize over all accepted
permutations $\Sigma$ and write
\begin{equation}
  P(\{ \hat\theta_i \} \mid \{ \theta_p \}) = \sum_{\sigma\in\Sigma}
  P(\{ \hat\theta_i \} \mid \{ \theta_{\sigma(i)} \}, \sigma) P( \sigma ) = 
  \frac{1}{|\Sigma|} \sum_{\sigma\in\Sigma}
  \prod_{i=1}^I P(\hat\theta_i \mid \theta_{\sigma(i)}) \; .
\end{equation}
Here we have assumed that all permutations have the same probability,
$P(\sigma) = 1/|\Sigma|$ where $|\Sigma|$ is the number of allowed
permutations: that is, $|\Sigma| = P^I$ for permutations with repetitions,
or $|\Sigma| = P! / (P - I)!$ for $I$-permutations (without repetitions).

Computationally, this expression can be evaluated in two different ways
depending on the permutation type. In the case of permutations with
repetitions we simply have
\begin{equation}
  P(\{ \hat\theta_i \} \mid \{ \theta_p \}) = \frac{1}{|\Sigma|}
  \prod_{i=1}^I \sum_{p=1}^P P(\hat\theta_i \mid \theta_p) \; .
\end{equation}
This expression, written in terms of log-quantities, becomes
\begin{equation}
  \log P(\{ \hat\theta_i \} \mid \{ \theta_p \}) = 
  -\log|\Sigma| + \sum_{i=1}^I \LSE_{p=1}^P 
  \log P(\hat\theta_i \mid \theta_p) \; ,
\end{equation}
where $\LSE_p$ denotes the LogSumExp (see, e.g.,
\citealp{10.1093/imanum/draa038}) operation over $p$. Its gradient is given by
\begin{align}
  \frac{\partial \log P(\{ \hat\theta_i \} \mid \{ \theta_p \})}{\partial \beta} 
  = \sum_{i=1}^I \sum_{p=1}^P & \mathrm{softmax}_p\bigl[\log 
  P(\hat\theta_i \mid \theta_p) \bigr] \cdot {} \notag\\
  & A_i^{-T} \mathbf{\Theta}_i \bigl( \hat\theta_i - \theta_p \bigr)\; ,
\end{align}
where $\mathrm{softmax}$ is the softmax function:
\begin{equation}
  \mathrm{softmax}_p(x_p) = \frac{\exp x_p}{\sum_{p'} \exp x_{p'}} \; .
\end{equation}
In the case of permutations without repetitions, instead, we cannot perform
the simplification above, and we have therefore
\begin{equation}
  \log P(\{ \hat\theta_i \} \mid \{ \theta_p \}) = -\log |\Sigma| + 
  \LSE_\sigma \biggl[ \sum_{i=1}^I
  \log P(\hat\theta_i \mid \theta_{\sigma(i)}) \biggr] \; .
\end{equation}
The gradient of this expression is
\begin{align}
  \frac{\partial \log P(\{ \hat\theta_i \} \mid \{ \theta_p \})}{\partial \beta} 
  = {} & \mathrm{softmax}_\sigma \biggl[\sum_{i=1}^I \log P(\hat\theta_i \mid \theta_p) \biggr] 
  \cdot {} \notag\\
  & \sum_{i=1}^I A_i^{-T} \mathbf{\Theta}_i \bigl( \hat\theta_i - \theta_{\sigma(i)} \bigr)\; ,
\end{align}
Finally, we note that the use of the normalization $1 / |\Sigma|$ above
introduces, effectively, a factor that penalizes cases where we have many
images predicted and only a few images observed. This is very sensible and
shows that the approach followed here naturally encapsulates a penalty factor
for unobserved images.

\subsection{Magnitudes}

We now consider the case of luminous image measurements and provide analytic
expressions for the (logarithmic) likelihood associated with the magnitude
measurements. We stress that, similarly to other source properties described
below, the use of individual measurements, instead of derived quantities (such
as, for example, the flux ratios) allows us to work with statistically
independent quantities. We note also that since the Jacobian of the lens
mapping is already used to compute the magnification factor (lens modules), we
cannot provide any derivative of the likelihood for magnitudes without higher
order derivatives of the lens mapping.

If we suppose that the intrinsic (unlensed) magnitude $M$ is a given
parameter, the generalization of the results of the previous subsection is
trivial: essentially, we just need to consider additional terms associated
with the magnitude measurements. For example, the result obtained for the
direct match is
\begin{align}
  2 \log {} & P(\{ \hat\theta_i \}, \{ \hat m_i \} \mid \{ \theta_i \}, M) = 
  \sum_{i=1}^I \log \biggl| \frac{\mathbf{\Theta}_i}{2\pi} \biggr| 
  - (\hat\theta_i - \theta_i)^T \mathbf{\Theta}_i (\hat\theta_i - \theta_i) \notag\\
  & {} + \log \frac{\lambda_i}{2\pi} 
  - \lambda_i (\hat m_i + \mathit{LM}_i - M)^2 \; .
\end{align}
The other methods are easily generalized.

The fact that the source magnitude $M$ enters the log-likelihood in a simple
way allows us to perform an important simplification and to avoid the use of
an explicit parameter for this quantity. We defer this discussion to
Sect.~\ref{sec:marginalized_source_parameters}.

\subsection{Quadrupole moments}

Sources with quadrupole measurements have a log-likelihood of the form
\begin{align}
  2 \log {} & P(\{ \hat\theta_i \}, \{ \hat m_i \}, \{ \hat Q_i \} \mid 
  \{ \theta_i \}, M, \mathbf{S}, \{ \nu_i \} ) = {} \notag\\
  \sum_{i=1}^I \biggl[ & \!\log \biggl| \frac{\mathbf{\Theta}_i}{2\pi} \biggr| - 
  (\hat\theta_i - \theta_i)^T \mathbf{\Theta}_i (\hat\theta_i - \theta_i) \notag\\ 
  & {} + 
  \log \frac{\lambda_i}{2\pi} - \lambda_i (\hat m_i + \mathit{LM}_i - M)^2 \notag\\
  & {} +
  \frac{\nu_i}{2} \log |\nu_i A_i^{T} \mathbf{S} A_i| 
  - \log \bigl[ 4 \pi \Gamma(\nu_i - 1) \bigr] + \frac{\nu_i - 3}{2} \log |\hat Q_i| 
  \notag\\ 
  & \qquad {} - \nu_i \mathrm{tr} (A^T_i \mathbf{S} A_i \hat Q_i) \biggr]
  \; ,
\end{align}
where we have called $\mathbf{S}$ the inverse of the source quadrupole and
where we have split in each line the contribution from the three different
measurements (positions, magnitudes, and quadrupole moments).

For this kind of source, we can apply arguments similar to the ones discussed
above. In particular, if we keep $M$ and $\mathbf{S}$ as free parameters, the
computation of the log-likelihood proceeds as for point-sources, with the
additional terms associated with the magnitudes and quadrupole moments.

\subsection{Time delay measurements}

For point-sources with associated time delays we can define a log-likelihood
function following closely what is done for luminous image measurements. The
equations, from a mathematical point of view, are identical, and we also have
again the possibility of performing a marginalization over the source time
$T$.

\section{Source-plane likelihood}
\label{sec:source_plane_likelihood}

\subsection{Positions}

The easiest way to evaluate the likelihood above is to avoid, as much as
possible, the computation of the predicted images $\{ \theta_i \}$. This
task requires the inversion of the nonlinear lens mapping, something
nontrivial and rather time consuming.

The approach uses a simple idea. Consider the lens equation
\begin{equation}
  \beta = f(\theta) \; .
\end{equation}
If $\beta$ is a regular value, this nonlinear equation admits local
inverses that we call $f^{-1}_i$: that is, each predicted image
$\theta_i$ of the $\beta$ is given by
\begin{equation}
  \theta_i = f^{-1}_i(\beta) \; .
\end{equation}
We can perform a Taylor expansion to first order of these equations around
$f(\hat\theta_i) \equiv \hat\beta_i$ to obtain
\begin{equation}
  \theta_i = f^{-1}_i(\beta) \approx f^{-1}_i(\hat\beta_i) + \left . 
  \frac{\partial f^{-1}_i}{\partial \beta} \right|_{\hat\beta_i} (\beta - \hat\beta_i) =
  \hat\theta_i + \left . \frac{\partial f^{-1}_i}{\partial \beta} \right|_{\hat\beta_i}
  (\beta - \hat\beta_i) \; .
\end{equation}
From the inverse function theorem, we can write the Jacobian of the inverse
$f_i^{-1}$ evaluated at $\hat\beta_i$ as the inverse of the Jacobian of
$f$ evaluated at the corresponding point, $\hat\theta_i$. We find
therefore
\begin{equation}
  \hat\theta_i - \theta_i = A_i^{-1} (\hat\beta_i - \beta) \; ,
\end{equation}
where we have called
\begin{equation}
  A_i = \left. \frac{\partial f}{\partial \theta} \right|_{\hat\theta_i} \; .
\end{equation}
Using this approximation in the log-likelihood above we find
\begin{equation}
  \log P(\{ \hat\theta_i \} \mid \beta) = \sum_{i=1}^I \frac{1}{2} \log{\left| 
  \frac{\mathbf\Theta_i}{2\pi} \right|} - \frac{1}{2} (\hat\beta_i - \beta)^T B_i
  (\hat\beta_i - \beta) \; ,
\end{equation}
where $B_i \equiv A_i^{-T} \mathbf\Theta_i A_i^{-1}$ is a symmetric matrix
representing the $i$-th precision $\mathbf\Theta_i$ mapped into the source
plane.

Taken as a function for $\beta$, this expression can be written as a
bivariate Gaussian:
\begin{equation}
  \log P(\{ \hat\theta_i \} \mid \beta) = \mathrm{const} - 
  \frac{1}{2} (\beta - \bar\beta)^T B (\beta - \bar\beta) \; ,
\end{equation}
where the mean $\bar\beta$ and the precision $B$ are given by
\begin{align}
  \label{eq:mean_precision}
  B & {} = \sum_{i=1}^I A_i^{-T} \mathbf\Theta_i A_i^{-1} = \sum_{i=1}^I B_i \; , \\
  \label{eq:best_source_position}
  \bar\beta & {} = B^{-1} \sum_{i=1}^I A_i^{-T} \mathbf\Theta_i A_i^{-1} \hat\beta_i = 
  B^{-1} \sum_{i=1}^I B_i \hat\beta_i \; .
\end{align}
As a result, we can immediately write the maximum-likelihood solution
(and its precision matrix) as $\bar\beta$ (and $B$).

In a frequentist approach, we could just substitute the maximum-likelihood
solution $\beta = \bar\beta$ into the likelihood and compute its
maximum-likelihood value: this, in turn, could be used in an outer
optimization loop, where we would change the lensing parameters until we
obtain a maximum of this source-plane likelihood.

If, instead, we adopt a Bayesian approach, we should proceed as explained
above and compute the marginalized likelihood over the source positions. Since
the conjugate prior in our case is also a bi-variate normal distribution, we
assume that $\beta$ is distributed as
\begin{equation}
  P(\beta \mid \hat\beta_0, B_0) = \sqrt{\biggl| \frac{B_0}{2\pi} \biggr|}
  \exp\left[ -\frac{1}{2} (\beta - \hat\beta_0)^T B_0 (\beta - \hat\beta_0) 
  \right] \; .
\end{equation}
Here $\hat\beta_0$ and $B_0$ are meta-parameters that describe the prior on
the source position: $\hat\beta_0$ is the center of the distribution, and
$B_0$ is its precision. Typically, one starts with a very loose prior, and
therefore one might set $B_0$ might be chosen as a (very) small multiple of
the identity matrix.

In this approach, we apply Bayes' theorem and write the posterior distribution
for $\beta$ as
\begin{equation}
  P(\beta \mid \{ \hat\theta_i \}) = \frac{\log P(\{ \hat\theta_i \} \mid \beta)
  P(\beta \mid \hat\beta_0, B_0)} {\int \log P(\{ \hat\theta_i \} \mid \beta')
  P(\beta' \mid \hat\beta_0, B_0) \, \mathrm{d}\beta'} \; \cdot
\end{equation}
A simple calculation then shows that the numerator of the posterior, the
expression $P(\{ \hat\theta_i \} \mid \beta) P(\beta \mid \beta_0, B_0)$,
taken as a function of $\beta$ is proportional to a multivariate normal
distribution with updated meta-parameters
\begin{align}
  B & {} = B_0 + \sum_{i=1}^I B_i = \sum_{i=0}^I B_i \; , \\
  \bar\beta & {} = B^{-1} (B_0 \beta_0 + B \beta) = B^{-1} \sum_{i=0}^I B_i \hat\beta_i \; .
\end{align}
We note how these results are formally identical to the ones found above, with
the (slight) difference that the sums start at $i = 0$ and includes,
therefore, the prior meta-parameters.

The normalizing constant, that is, the denominator in the posterior
distribution, is generally called the evidence or the marginalized likelihood.
Its logarithm can be written as
\begin{equation}
  2 \log E = \sum_{i=0}^I \biggl[ \log{\biggl| \frac{\mathbf\Theta_i}{2\pi}\biggr| } - 
  \hat\beta_i^T B_i \hat\beta_i \biggr] - 
  \biggl[ \log \biggl| \frac{B}{2\pi} \biggr| - \bar\beta^T B_0 \bar\beta \biggr]
  \; ,
\end{equation}
or, equivalently,
\begin{equation}
  2 \log E = \sum_{i=0}^I \biggl[ \log{\biggl| \frac{\mathbf\Theta_i}{2\pi}\biggr| } - 
  (\hat\beta_i - \bar\beta)^T B_i (\hat\beta_i - \bar\beta)
  \biggr] - \log{\biggl| \frac{B}{2\pi} \biggr|} \; .
\end{equation}
We note that in these expressions we have defined $\mathbf\Theta_0 \equiv B_0$,
so to be able to include the prior normalizing constant in a way similar to
the other measurements. The evidence $E$, in our Bayesian approach, replaces
the likelihood, as it is effectively a likelihood marginalized over the source
position (or, more generally, the source parameters).

In many cases, it is convenient to consider also an evidence based on a flat
prior. In our notation, this corresponds to taking the limit
\begin{equation}
  E_0 = \lim_{\mathbf\Theta_0 \rightarrow 0} E \, \biggl| 
  \frac{\mathbf\Theta_0}{2\pi} \biggr|^{-1/2} \; ,
\end{equation}
that is, to the use of an improper prior $P(\beta_0) = 1$. The expressions
obtained for $\log E_0$ are formally identical to the expressions above for
$\log E$, with the sums starting at $i = 1$. More explicitly:
\begin{equation}
  2 \log E_0 = \sum_{i=1}^I \biggl[ \log{\bigg| \frac{\mathbf\Theta_i}{2\pi}\biggr| } - 
  \hat\beta_i^T A_i^{-T} \mathbf\Theta_i A_i^{-1} \hat\beta_i \biggr] - 
  \log{\biggl| \frac{B}{2\pi} \biggr|} + \bar\beta^T B \bar\beta \; ,
\end{equation}
or, equivalently,
\begin{equation}
  2 \log E_0 = \sum_{i=1}^I \biggl[ \log{\bigg| \frac{\mathbf\Theta_i}{2\pi}\biggr| } - 
  (\hat\beta_i - \bar\beta)^T A_i^{-T} \mathbf\Theta_i A_i^{-1} (\hat\beta_i - \bar\beta)
  \biggr] - \log{\biggl| \frac{B}{2\pi} \biggr|} \; .
\end{equation}
Naively, we could imagine that this expression is equivalent to a substitution
of the maximum-likelihood solution $\beta = \bar\beta$ inside the likelihood.
This, however, is not the case: the additional term to the right,
$\log|B/2\pi|$, is a nontrivial addition since it is influenced by the
jacobian matrices of the lens-mapping equation at the observed image positions
$\{ A_i \}$, and therefore depends on the chosen lens model.\footnote{The use
of this marginalized likelihood is controlled in the code by the flag
\texttt{bayesianfactor}.}

\subsection{Magnitudes}

Luminous sources are characterized by their position and luminosity. Here, we
prefer to describe them in terms of magnitudes, as this is more directly
interpreted in terms of practical measurements, and because of the possible
use of a specific prior (see below).

As usual, we can proceed as before and assume initially that the unknown
source magnitude $M$ has a normal (Gaussian) distribution for its prior:
\begin{equation}
  P(M \mid \hat M_0, \lambda_0) = \sqrt{\frac{\lambda_0}{2\pi}} \exp \biggl[ -\frac{1}{2} 
  \lambda_0 (\hat M - M_0)^2 \biggr] \; .
\end{equation}
With a calculation similar to the one discussed above, we easily find that the
posterior probability distribution for the source magnitude $M$ is a
Gaussian with precision (inverse variance)
\begin{equation}
  \lambda = \sum_{i=0}^I \lambda_i
\end{equation}
and mean
\begin{equation}
  \bar M = \lambda^{-1} \sum_{i=0}^I \lambda_i (\hat m_i + \mathit{LM}_i) = 
  \lambda^{-1} \sum_{i=0}^I \lambda_i \hat M_i \; ,
\end{equation}
where we have called $\hat M_i \equiv \hat m_i + \mathit{LM}_i$.
The corresponding evidence is
\begin{equation}
  2 \log E = \sum_{i=0}^I \biggl[ \log \frac{\lambda_i}{2\pi} - 
  \lambda_i \hat M_i^2 \biggr] 
  - \biggl[ \log \frac{\lambda}{2\pi} - \lambda \bar M^2 \biggr] \; ,
\end{equation}
or, equivalently,
\begin{equation}
  \label{eq:logl_mag_sp}
  2 \log E = \sum_{i=0}^I \biggl[ \log \frac{\lambda_i}{2\pi} - 
  \lambda_i \bigl( \hat M_i - \bar M \bigr)^2 \biggr] 
  - \log \frac{\lambda}{2\pi} \; .
\end{equation}

As before, it can be sensible to consider the case of an improper flat prior
and consider the limit $\lambda_0 \rightarrow 0$. Again, this produces an
evidence $E_0$ which is formally identical to the expression for $E$, with the
sum starting at $i = 1$.

For magnitudes, it might also be interesting to consider the improper prior
\begin{equation}
  P(\hat M_0 \mid \alpha) \propto \mathrm{e}^{\alpha \hat M_0} \; .
\end{equation}
This model is justified by the number counts of distant sources, which under
mild hypotheses are distributed according to an exponential law with $\alpha =
0.6 \ln 10 \approx 1.38$. In this case, we find
\begin{align}
  \lambda & {} = \sum_{i=1}^I \lambda_i \; , \\
  \bar M & {} = \lambda^{-1} \biggl[ \alpha + \sum_{i=1}^I \lambda_i \hat M_i \biggr] \; ,
\end{align}
while the evidence under the new prior is still
\begin{equation}
  2 \log E_\alpha = \sum_{i=1}^I \biggl[ \log \frac{\lambda_i}{2\pi} - 
  \lambda_i \hat M_i^2 \biggr] 
  - \biggl[ \log \frac{\lambda}{2\pi} - \lambda \bar M^2 \biggr] \; .
\end{equation}
We note how $\alpha = 0$ returns the usual expressions for a flat improper
prior.

\begin{table*}
  \centering
  \caption{Operation schemes of the code, sorted by increasing computational
  complexity.}
  \label{tab:operating_schemes}
  \begin{tabular}{lcccl}
    \textbf{Operating scheme} & \textbf{Inversion of the} & 
    \textbf{Source} & \textbf{Other source} & \textbf{Notes} \\
    & \textbf{lens equation} & \textbf{position} & \textbf{parameters} & \\
    \hline
    Source plane & none & marginalized & marginalized & Cannot discover new images \\
    Fast image plane & required & marginalized & marginalized & Laplace's approximation \\
    Standard image plane & required & free & marginalized & Exact marginalization \\
    Full image plane & required & free & free & Rarely needed \\
  \end{tabular}
  \tablefoot{\Gravity\ allows the user to choose the operating scheme, and to
    take advantage of results obtained by a simpler scheme to speed up the
    inference process in a more complex one through the use of the reference
    distribution (see Sect.~\ref{sec:prior_initial_guesses}).}
\end{table*}

\subsection{Quadrupole measurements}

Let us call $\mathbf{S}$ the inverse of the source quadrupole. Since $\langle
\hat Q_i \rangle = \nu_i \mathbf{W}_i^{-1}$, and since quadrupole moments
transform as rank-two tensors, we have
\begin{equation}
  \mathbf{W}_i = \nu_i A_i^{T} \mathbf{S} A_i \; .
\end{equation}
If we insert this equation in the likelihood for the quadrupole moments we find
\begin{align}
  & \log P \bigl(\{ \hat Q_i \} \mid \mathbf{S}, \{ \nu_i \} \bigr) =
  \sum_{i=1}^I \biggl[ \frac{\nu_i}{2} \log \bigl| \nu_i A_i^{T} \mathbf{S} A_i \bigr| 
  - \log \bigl[ 4 \pi \Gamma(\nu_i - 1) \bigr] \notag\\
  & \qquad {} + \frac{\nu_i - 3}{2} \log \bigl| 
  \hat Q_i \bigr| - \frac{\nu_i}{2} \mathrm{tr} (A^T_i \mathbf{S} A_i \hat Q_i) 
  \biggr] \; .
\end{align}

To proceed, we now consider the prior distribution for the inverse source
quadrupole $\mathbf{S}$. We decide to use for this quantity also as a Wishart
prior, with meta-parameters $\hat O_0$ and $\nu_0$:
\begin{align}
  & \log P(\mathbf{S} \mid \hat O_0, \nu_0) = \frac{\nu_0}{2} \log \bigl| \nu_0 \hat O_0 \bigr| - 
  \log \bigl[ 4 \pi \Gamma(\nu_0 - 1) \bigr] \notag\\
  & \qquad {} + \frac{\nu_0 - 3}{2} \log | \mathbf{S} |
  - \frac{\nu_0}{2} \mathrm{tr}(\hat O_0 \mathbf{S}) \; .
\end{align}
We note that our choice for the prior of $S$ implies that the source quadrupole
has an inverse Wishart distribution \citep{10.1093/biomet/20A.1-2.32} as
prior. The product between the prior and the likelihood, taken as a function
of $\mathbf{S}$, still has the same form as the prior. To show this, one can
use the cyclic property of the trace and the expression for the determinant of
the product of matrices. The posterior has updated meta-parameters
\begin{equation}
  \nu = \nu_0 + \sum_{i=1}^I \nu_i = \sum_{i=0}^I \nu_i
\end{equation}
and
\begin{equation}
  \bar O = \frac{1}{\nu} \biggl[ \nu_0 \hat O_0 + 
  \sum_{i=1}^I \nu_i A_i \hat Q_i A_i^T  \biggr] = 
  \frac{1}{\nu} \sum_{i=0}^I \nu_i \hat O_i \; ,
\end{equation}
where we have called $\hat O_i \equiv A_i \hat Q_i A_i^T$. The expression
above shows that, essentially, the parameter $\bar O$, representing an
estimate for the source quadrupole, is a weighted average of the measured
quadrupole moments projected into the source plane, with weights $\nu_i$.

The evidence associated with the updated distribution is
\begin{align}
  \label{eq:logl_quad_sp}
  \log E = {} & \sum_{i=0}^I \frac{\nu_i}{2} \log \bigl| \nu_i \hat O_i \bigr| - 
  \frac{\nu}{2} \log \bigl| \nu \bar O \bigr| \notag \\ 
  & {} - 
  \sum_{i=0}^I \log \bigl[ 4 \pi \Gamma(\nu_i - 1) \bigr] +
  \log \bigl[ 4 \pi \Gamma(\nu - 1) \bigr] \notag\\
  & {} - 
  \frac{3}{2} \sum_{i=1}^I \log \bigl| \hat Q_i \bigr| \; .
\end{align}
This expression, whose symmetry is evident, depends on the lensing model only
through the first two terms, as are both functions of the set of lensing
Jacobian matrices $\{ A_i \}$. The other terms, instead, only depend on the
measurements. We note also how the largest evidence is obtained when all
projected quadrupole moments $\{ \hat O_i \}$ agree, as in this case there
are no cancellation effects in the expression for $\bar O$.

As before, we can consider an uninformative improper prior characterized by
$\nu_0 = 0$. In that case, the expressions above are just essentially the
same, with all sums starting from $i = 1$.

\subsection{Time delays}

As mentioned already, the equations that describe the time delays are formally
identical to the ones that describe the magnitudes. They are both scalar
quantities that depend on a single scalar source parameter (the unlensed
magnitude $M$ or the reference time $T$) through additive terms (the
lensing modules $\mathit{LM}_i$ or the time delays $T_i$). Moreover, for
the measurements of both quantities, we assume simple normal distributions.

Therefore, all equations found in the previous section apply, with the due
variable changes, for time delay measurements. In particular, with a flat
prior for $T$ we find
\begin{align}
  \tau & {} = \sum_{i=1}^I \tau_i \; , \\
  \bar T & {} = \tau^{-1} \sum_{i=1}^I \tau_i (\hat t_i - T_i) \; ,
\end{align}
and evidence
\begin{equation}
  \label{eq:logl_time_sp}
  2 \log E_0 = \sum_{i=1}^I \biggl[ \log \frac{\tau_i}{2\pi} - 
  \tau_i \bigl( \hat t_i - T_i - \bar T \bigr)^2 \biggr] 
  - \log \frac{\tau}{2\pi} \; .
\end{equation}

\section{Image-plane likelihood optimizations}
\label{sec:image_plane_likelihood_optimizations}

\subsection{Marginalized source parameters}
\label{sec:marginalized_source_parameters}

The source-plane analysis is a powerful tool for analyzing lensing systems, as
it allows us to study a lensing system relatively quickly for two main
reasons: (1) we do not need to invert the lens mapping to compute the
predicted images and (2) we do not need to infer over the source parameters,
as these are already marginalized over. This is particularly useful when the
source parameters are not of interest, which is often the case.

However, the source-plane analysis has also a possible drawback: it is
essentially based on a local linearization of the lens mapping, and therefore
it might not be sufficiently accurate in some situations. Additionally, it
would not highlight possible issues of the lensing model, as the predictions
of bright counter-images that are not observed in the data.

In these cases, it might be necessary to move to an image-plane analysis. In
doing so, however, we can still take advantage of the fact that source
parameters such as the source magnitude or the source reference time $T$
appear quadratically in the log-likelihood. This allows us to perform an exact
marginalization over these parameters, and to retain the source-plane
likelihood only the position of each source. Specifically, the likelihood is
composed of a term associated with the position, identical to
Eq.~\eqref{eq:logl_pos_ip}, and to other terms associated with the other
source parameters, as in Eqs.~\eqref{eq:logl_mag_sp}, \eqref{eq:logl_quad_sp},
or \eqref{eq:logl_time_sp}, depending on the source type. The resulting
likelihood can, as usual, be computed for the various match methods (direct
association, best match, or marginalized matches). Of course, in case the
source redshift is unknown, it is included as an additional free parameter.

We call this approach the standard image-plane analysis, to distinguish it
from the full image-plane analysis, performed by including all source
parameters in the inference. We stress that in the standard image-plane
analysis there are no approximations involved.

\subsection{Laplace's method}

A further optimization can be obtained by using Laplace's approximation
\citep[see][]{MacKay} to avoid altogether the use of source parameters. This
approach is based on a local quadratic approximation of the log-posterior
around its maximum, and is therefore equivalent to assuming a normal
distribution for the log-posterior. In this case, one can analytically compute
the marginalization of the log-posterior over all source parameters, including
the source position.

A full implementation of this method requires the computation of the local
maximum of the posterior, together with its Hessian matrix. The latter is
relatively easy to compute, as one can resort to automatic differentiation to
compute the derivatives of the log-posterior. The former, instead, is more
difficult, as it requires the solution of a nonlinear equation, with all the
related issues (non-locality of the solution, convergence, etc.).

In practice, we have found that a good compromise is obtained by applying
Laplace's method with the following prescriptions:
\begin{itemize}
  \item The maximum is computed as in Eq.~\eqref{eq:best_source_position}
  by taking a weighted average of the observed image positions (together with
  their uncertainties) mapped into the source plane;
\item The inverse of the Hessian matrix is computed as
Eq.~\eqref{eq:mean_precision} by mapping the precision matrix of the observed
image positions into the source plane.
\end{itemize}

We note that, in this approach, that we name the ``fast image-plane
analysis,'' we still invert the lens mapping to compute the various predicted
images associated with a given source, and to compute from their positions the
image-plane likelihood. However, we gain in terms of computational speed, as
we do not need to include the source position as additional free parameters in
the optimization process. A summary of the possible operating schemes of the
code is reported synthetically in Table~\ref{tab:operating_schemes}.

\section{Technical aspects}
\label{sec:technical_aspects}

During the implementation of \Gravity, particular care has been devoted to
the numerical stability and efficiency of the code. As a result, the code is
able to handle hundreds of images and lensing parameters, and to efficiently
explore the parameter space. To test the performance of the code, we
reconsidered a lensing system that has been studied in the literature, the
``Pandora'' cluster Abell~2744 \citep{2023ApJ...952...84B}. For this system we
used exactly the same data and exactly the same lens model as in the original
paper, so that we could both validate the code and compare its performance
with the one of LensTool \citep{1996ApJ...471..643K, 2007NJPh....9..447J}, a
widely used lensing code also based on the Bayesian approach. The model
includes 181 halos (with 9 of them individually parametrized), 149 images
divided into 50 families at more than 20 different redshifts, and 8 sources with
unknown redshifts (so that their redshifts are additional free parameters); in
total, the model has 50 free parameters.

The source-plane analysis of this system in \Gravity\ takes approximately 30
minutes on a 32-core workstation, while the fast image-plane analysis takes
approximately 12 hours. The same analysis in LensTool takes approximately 3
weeks on a similar workstation (using, however, 100-core instead of 32,
P.~Bergamini, private communication). We estimate therefore a gain of a factor
of $\sim 100$ in terms of computational time. This is a remarkable result, and
it opens new opportunities to study lensing systems at a level of complexity
and detail that was not possible before. Below we highlight some of the
technical aspects that allowed us to obtain this result.

\subsection{Code tailoring and optimization}

\Gravity\ is implemented in the Julia language \citep{bezanson2017julia}, and
therefore it benefits from the high performance of this language and of the
LLVM compiler infrastructure. In particular, it takes advantage of the
just-in-time compilation of the code to produce a highly optimized version of
the likelihood computation, tailored to the specific lensing system under
consideration. For example, \Gravity\ automatically detects if there
is a need to update the cosmological distances (because either the redshifts or
the cosmological model includes free parameters), or, in the case of multiplane
lensing, if it is necessary to project the observed positions of the
background lenses to obtain their real positions in the background lensing
plane.

Particular care has been devoted to some technical aspects of the Julia
implementation, such as the use of type-stable code, which allows the compiler
to produce machine code that is as fast as C or Fortran code (avoiding one of
the penalties of dynamically typed languages). Additionally, throughout the
development, we used micro-benchmarking techniques for critical parts of the
code (such as the computation of the lens mapping, or the inversion of the
lens mapping) to optimize the various algorithms used. The same
micro-benchmarking techniques are also used for some tasks at runtime to
select the most efficient algorithm among a set of equivalent ones.

\subsection{Parallelization}

The code uses the parallelization capabilities of the Julia language to
distribute the computation of the likelihood over multiple cores and,
depending on the sampling technique used, also over distributed systems. The
code also makes use of single-instruction multiple-data (SIMD) instructions when
available, to further speed up the computation even in the case of a single
core; to this goal, the code uses also the SLEEF library \citep{SLEEF}.

\subsection{Automatic differentiation}

The code uses the automatic differentiation capabilities of the Julia language
to compute the derivatives of the log-posterior with respect to the lensing
parameters. Currently, the code uses forward-mode automatic
differentiation implemented in the \texttt{ForwardDiff.jl} package
\citep{RevelsLubinPapamarkou2016}. This allows the use of sampling techniques
based on the gradient of the log-posterior, such as Hamiltonian Monte Carlo
(see below Sect.~\ref{sec:sampling_techniques}).

\begin{figure}
  \centering
  \includegraphics[width=\hsize]{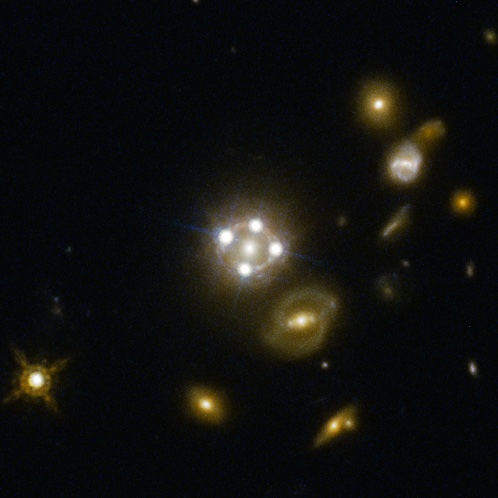}
\caption{Lensing system HE0435$-$1123. As shown by this image,
the lensing galaxy is a massive elliptical galaxy, surrounded by a group of
galaxies. The barred spiral galaxy below the lensing galaxy is in the
background. Credits: ESA/Hubble, NASA, Suyu et al.}
  \label{fig:he0435}
\end{figure}

\subsection{Sampling techniques}
\label{sec:sampling_techniques}

The code allows for the use of different sampling techniques:
\begin{description}
  \item[Metropolis-Hastings.] The vanilla Metropolis-Hastings algorithm
  \citep{Metropolis, Hastings} is the simplest sampling technique. It requires
  careful tuning of the proposal distribution, and it is generally not very
  efficient.
  \item[Affine invariant ensemble sampler.] A very efficient algorithm based on
  the ensemble sampler of \citet{2010CAMCS...5...65G} (and known in the Python
  community through the library \texttt{emcee}). This algorithm is very
  efficient in exploring the parameter space when the posterior distribution is
  unimodal, which is often the case.
  \item[Elliptical slice sampling.] Uses the algorithm proposed in
  \citet{pmlr-v9-murray10a}.
  \item[Hamiltonian Monte Carlo.] The Hamiltonian Monte Carlo (HMC) algorithm is
  a more sophisticated sampling technique (\citealp{2011hmcm.book..113N}; see
  also \citealp{2017arXiv170102434B}). It is based on the introduction of a
  fictitious momentum variable, which allows for a more efficient exploration of
  the parameter space. The use of this algorithm requires the derivative of the
  log-posterior, which in our code is computed using automatic differentiation.
  The library used in \Gravity, \citet{xu2020advancedhmc}, also implements the
  No-U-Turn Sampler (NUTS) approach of \citet{2011arXiv1111.4246H}.
  \item[Metropolis-adjusted Langevin algorithm.] This method, called informally
  MALA, can be seen as an HMC algorithm with a single leap-frog time step
  \citep{10.1063/1.436415}.
  \item[Parallel tempering.] The parallel tempering algorithm is a
  generalization of the Metropolis-Hastings algorithm
  \citep{1986PhRvL..57.2607S,1992EL.....19..451M}. In this algorithm, several
  chains are run in parallel, each at a different ``temperature'' $\beta$,
  that is, with a modified likelihood function $P(\mathbf{D} \mid \mathbf{M},
  \beta) = P(\mathbf{D} \mid \mathbf{M})^\beta$. When $\beta = 0$, the
  likelihood is flat, and the chain explores the prior distribution, while when
$\beta = 1$ the chain explores the posterior distribution. The algorithm
  alternates between a local exploration phase, when the chains are run
  independently, and a global exploration phase, when the chains are
  coupled and the states are exchanged between them. This algorithm is
  particularly efficient in exploring the parameter space when the posterior
  distribution is multimodal. 
  \item[Nested sampling.] The nested sampling algorithm
  \citep{2004AIPC..735..395S, 2008arXiv0801.3887C} is a different approach to
  Bayesian inference. It is based on the idea of transforming the integral over
  the parameter space into a sum of integrals over the likelihood, and then
  sampling from the likelihood. The algorithm is particularly efficient in
  computing the evidence, and it is particularly useful when the posterior
  distribution is multimodal. The code uses the implementation of
  \citet{miles_lucas_2021_5808196}.
\end{description}

For completeness, \Gravity\ also includes robust methods to obtain the maximum
likelihood or the maximum-a-posteriori solution, based on
globally convergent optimization algorithms \citep{metaheuristics2022}.

\subsection{Unit testing}

The code includes an extensive suite of unit tests, which are run at each
commit to the code repository. The tests cover all the critical parts of the
code, such as the computation of the lens mapping, the grid algorithm, the
computation of the likelihood, and the sampling techniques. The tests also
include a coverage analysis, which helps us to identify parts of the code that
require further testing.

\section{HE0435$-$1123}
\label{sec:application}

\begin{figure}
  \centering
  \includegraphics[width=\hsize]{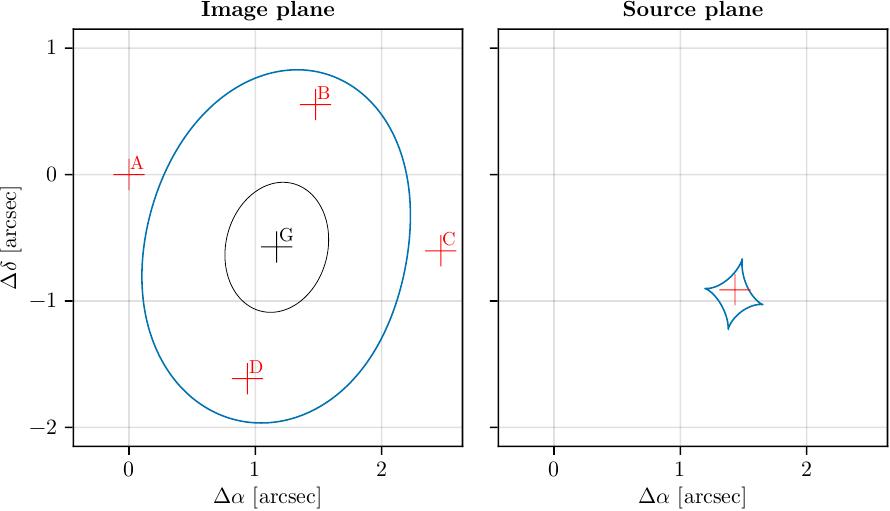}
\caption{Best-fit lensing model. \textbf{Left.} Observed image positions (red
crosses) and the critical curves (blue line) for the lensing system
HE0435$-$1123 in the singular-isothermal ellipsoid model. The black ellipse
indicates the axis ratio and orientation of the main lens. \textbf{Right.} The
predicted source position (red cross) and the caustic lines.}
  \label{fig:he0435_lens}
\end{figure}

\begin{figure*}
  \centering
  \includegraphics[width=\hsize]{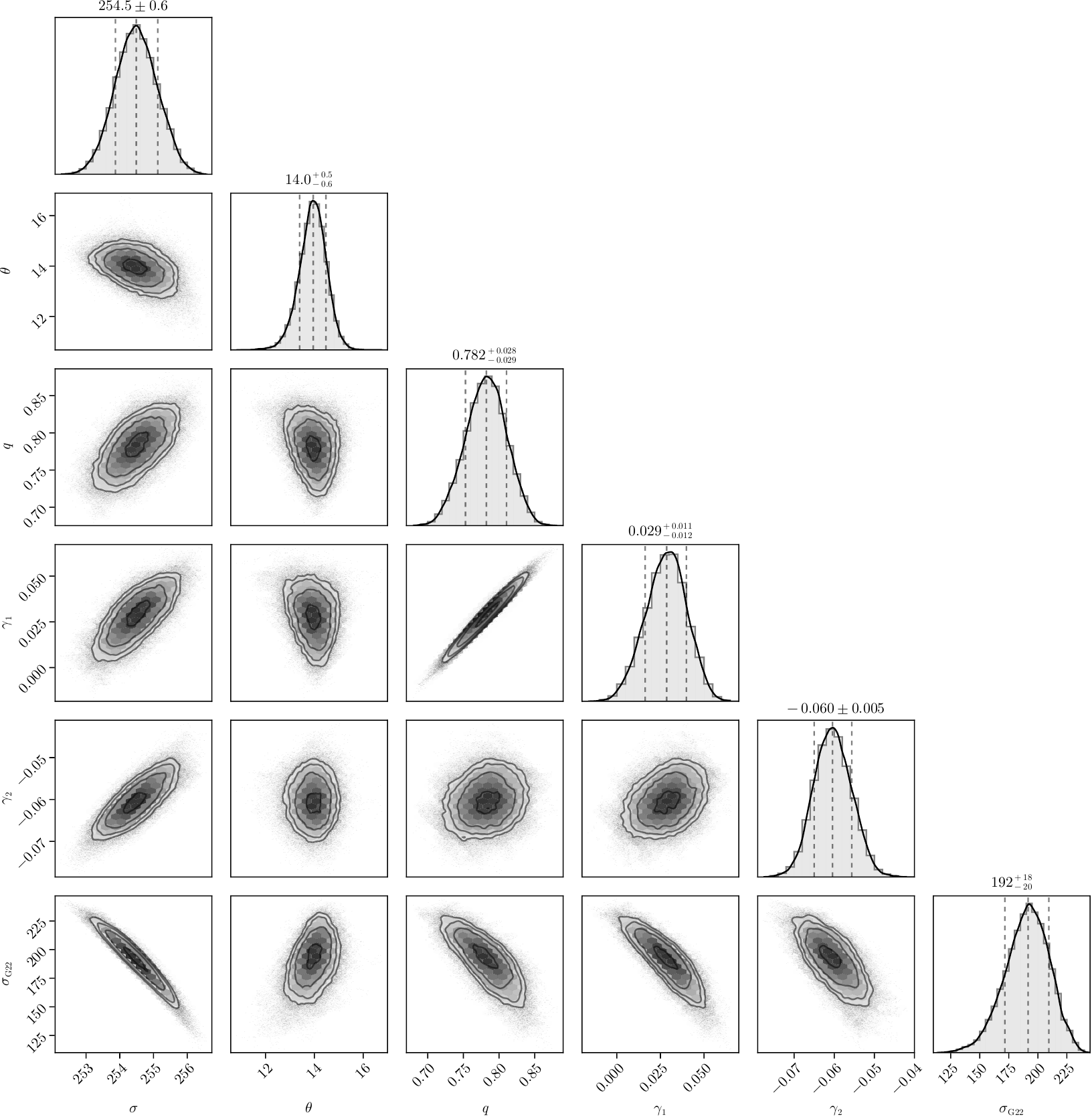}
\caption{Joined distribution of the lensing parameters for the system
HE0435$-$1123 in a model using a singular isothermal ellipsoid as the main
lens. The contours enclose 68\% and 95\% of the probability.}
  \label{fig:he0435_pairplot_nie}
\end{figure*}

\begin{figure*}
  \centering
  \includegraphics[width=\hsize]{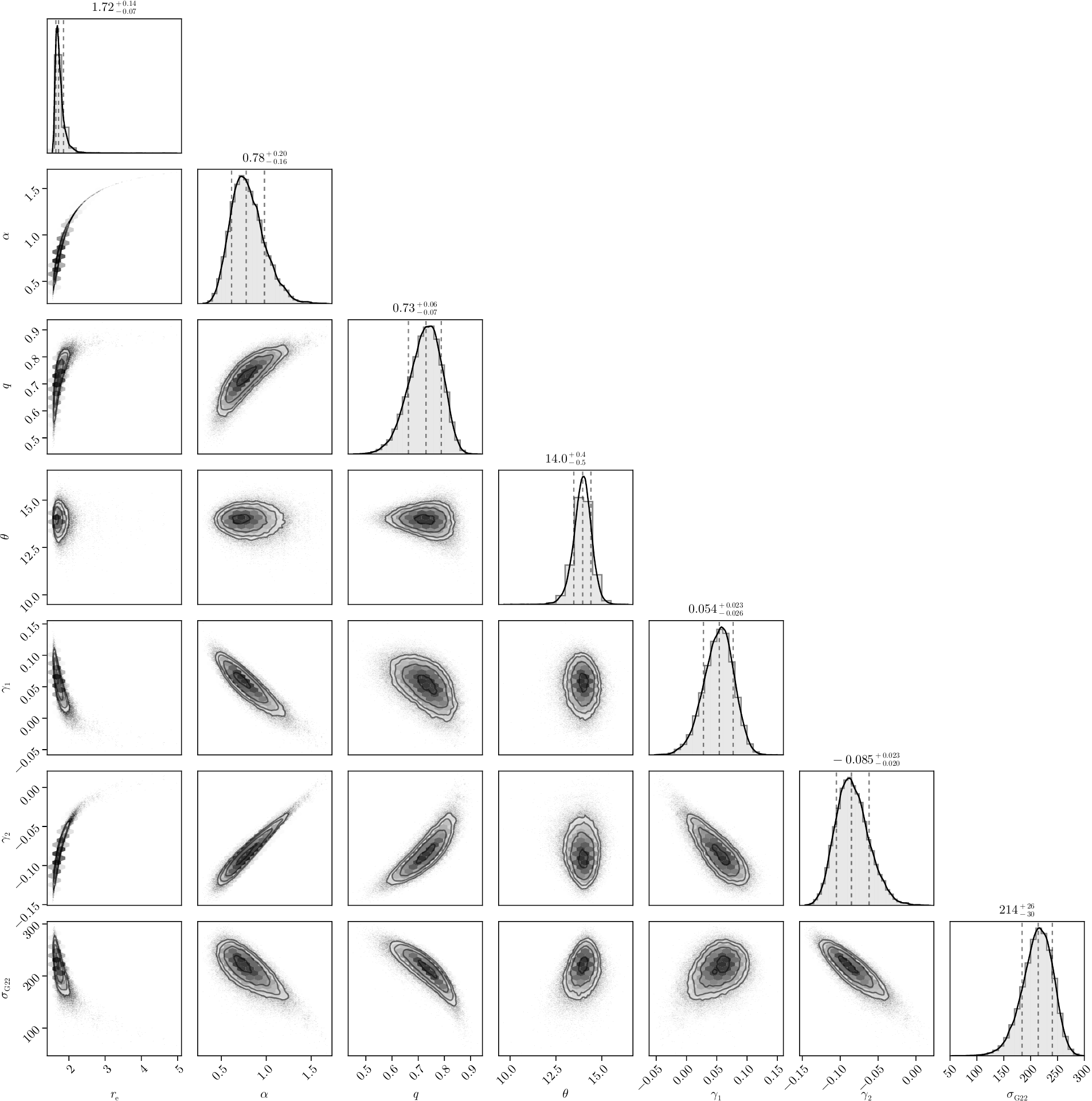}
  \caption{Same as Fig.~\ref{fig:he0435_pairplot_nie}, but using a singular
  power-law mass profile for the main lens. The contours enclose 68\% and 95\%
  of the probability.}
  \label{fig:he0435_pairplot_pow}
\end{figure*}

As a test application for a galaxy-scale lens, we consider the lensing system
HE0435$-$1123. This is a very well-known quadruply imaged quasar at redshift
$z_s = 1.69$ \citep{2000A&A...358...77W} lensed by a foreground galaxy at
redshift $z_l = 0.454$ \citep{2005AJ....129.2531M}. The system has been the
subject of several studies, often focusing on the time delays between the
images \citep{2006A&A...451..759E,2006ApJ...640...47K,2017MNRAS.465.4895W}. 

The lensing galaxy is a massive elliptical galaxy, with an Einstein radius of
$\theta_E \approx 1.61$ arcsec, but the whole system is part of a group of
galaxies. The lensing system is shown in Fig.~\ref{fig:he0435_lens}. Wide
Field Camera 3 (WFC3) grism observations have been used to show that the main
lensing galaxy lacks significant substructures and its gravitational potential
can be well described by an isothermal ellipsoid \citep{2017MNRAS.471.2224N}.
Therefore, in our first model, the main lensing galaxy is taken to have a
singular isothermal ellipsoid mass profile (SIE). The model also includes a
second lens to account for the closest perturber (a spiral galaxy in the
background of the main lens, at $z = 0.7821$), and a shear to account for the
external field.

Since the perturber galaxy is in the background of the main lens, we need to
use the full multiplane lensing equations to describe the system. We note also
that in doing so, we take into account the fact that the observed position of
the perturbing galaxy differs from the real position because of the
gravitational lensing effect of the main galaxy.

Our lensing model includes therefore 6 parameters associated with the lensing
system (which we list together with their flat prior distributions): the main
lens velocity dispersion (100 .. \SI{500}{km/s}), the lens axis ratio (0.25 ..
1) and position angle ($-90^\circ$ .. $90^\circ$), the external shear (-0.3 ..
0.3 in both orientations), and the perturbing galaxy velocity dispersion (50
.. \SI{250}{km/s}). The source has been modeled as a luminous point-like
object, with large normal prior distributions for all parameters (the position
coordinates and the magnitude). The constraints include the positions and
magnitudes for the four observed point-like images; the associated data and
uncertainties are taken from \citet{2005AJ....129.2531M}. For the magnitudes,
we have used larger errors (0.1 magnitudes) to include the potential
presence of microlensing effects and unaccounted luminosity variations
associated with the different time delays of the various images. The model has
therefore 9 free parameters (6 for the lens and 3 for the source) and 12
observational constraints.

For this system, as well as for other ones considered during the development
of \Gravity, we have compared many relevant lensing properties (such as the
deflection angle, the magnification, or the lensing potential) obtained with
our code and with alternative programs. We have made sure that if one performs
the appropriate transformations, the results obtained with different programs
(including LensTool, \citealp{1996ApJ...471..643K}, and Glafic,
\citealp{2010PASJ...62.1017O}) are consistent, to the level of numerical
precision, to each other.

Figure~\ref{fig:he0435_pairplot_nie} shows the combined probability distributions
for all lens parameters, obtained from a Bayesian inference in the imageplane.
The log-evidence of the model has been estimated to be $\ln Z \approx -24.2$.
These results have been obtained in less than one hour on a laptop.

For comparison, we have also considered an alternative model, where the main
lens is modeled as a singular power-law ellipsoid (implemented in \Gravity
through the fast algorithm of \citealp{2015A&A...580A..79T}). The posterior
distribution for the various parameters is shown in
Fig.~\ref{fig:he0435_pairplot_pow}. The first two parameters shown,
$r_\mathrm{e}$ and $\alpha$, represent a generalization of the lens Einstein
radius for elliptical lenses (computed from a circularized version of the mass
profile) and the slope of the power-law profile, $\kappa(x) \propto
x^{\alpha - 2}$.

We note how the slope of the power-law profile is poorly constrained from
these data alone, and is anyway essentially consistent with an isothermal one
$\alpha = 1$. We note also how some parameters present rather nasty
degeneracies: in particular, the Einstein radius of the main lens is strongly
correlated to the slope $\alpha$. Curved degeneracies such as this one are
particularly complicated to sample (and are in fact often used to test the
performance of optimization algorithms, see, e.g.,
\citealp{10.1093/comjnl/3.3.175}), but the code can handle them efficiently.

The log-evidence associated with this model is $\ln Z \approx -24.7$. These
results give very weak evidence in favor of the isothermal profile.

\section{Conclusions}
\label{sec:conclusions}

We have presented \Gravity, a new Julia-based code for the analysis of strong
gravitational lensing systems. The code is based on a Bayesian approach, and
it allows for the inference of the lensing parameters of a system, together
with the marginalization of the source parameters. The code is highly parallel
and able to handle hundreds of images and lensing parameters and still
efficiently explore the parameter space.

The algorithms implemented in \Gravity\ are all based on sound statistical
frameworks, obtained through the use of the marginalization over nuisance
parameters, in contrast to the use of optimizations. This allows for a robust
and efficient exploration of the parameter space, and for a reliable
estimation of the uncertainties associated with the lensing parameters.

In the future papers of this series, we will focus on the analysis of extended
sources, for which we will consider both parametric and non-parametric models.
We will also consider a combined weak + strong lensing analysis, carried out in a
way that is consistent with the Bayesian approach of the code.

\begin{acknowledgements}
  This work has benefited from many tests, discussions, and interactions with
  several collaborators, including Davide Abriola, Pietro Bergamini, Claudio
  Grillo, Massimo Meneghetti, and Piero Rosati. We are grateful to them for
  their help and support.
\end{acknowledgements}

\bibliographystyle{aa} 
\bibliography{references}

\begin{appendix}
\section{YAML configuration file}

\def\red#1{\textcolor{red}{#1}}%
\def\blue#1{\textcolor{blue}{#1}}%
\def\teal#1{\textcolor{teal}{#1}}%
\def\violet#1{\textcolor{violet}{#1}}%
\def\plusminus{$\mathtt{\pm}$}%

\begin{Verbatim}[fontsize=\small,commandchars=\\\{\}]
\red{cosmology}: 
  Omega_m: \blue{0.3}
  h_0: \blue{0.7}
  Omega_r: \blue{0.0}
  w_0: \blue{-1}
\red{sampling}:
  scheme: \blue{sourceplane}
  \violet{likelihood-options}:
    bayesianfactor: \blue{true}
    matches: \blue{all}
    duplicates: \blue{true}
    missedpredictionspenalty: \blue{true}
    fitsource: \blue{true}
  algorithm: \blue{hmc}
  \violet{algorithm-options}:
    method: \blue{NUTS}
    warmup: \blue{5000}
    iterations: \blue{50000}
  threads: \blue{auto}
  random-seed: \blue{1}
\red{lenses}:
  - \violet{NIE}:
    name: \blue{main}
    z: \blue{z_lens}
    x: \blue{x_lens}
    sigma: \blue{100} \teal{..} \blue{500}
    s: \blue{0.0}
    theta: \blue{-pi/2} \teal{..} \blue{pi/2}
    q: \blue{0.25} \teal{..} \blue{1}
- \violet{Shear}:
    name: \blue{shear}
    z: \blue{z_lens}
    x: \blue{x_lens}
    gamma: \blue{(-0.3, -0.3)} \teal{..} \blue{(0.3, 0.3)}
- \violet{SIS}:
    name: \blue{G22}
    z: \blue{z_G22}
    x: \blue{x_G22}
    sigma: \blue{50} \teal{..} \blue{250}
\red{sources}:
  - \violet{Luminous}:
    z: \blue{1.689}
    x: \blue{(0.0, 0.0)} \teal{\plusminus} \blue{10.0}
    mag: \blue{20.0} \teal{\plusminus} \blue{10.0}
    \red{images}:
      - \violet{Luminous}:
        x: \blue{(0.0, 0.0)} \teal{\plusminus} \blue{0.002}
        mag: \blue{18.545} \teal{\plusminus} \blue{0.1}
      - \violet{Luminous}:
        x: \blue{(-2.4687, -0.6033)} \teal{\plusminus} \blue{0.002}
        mag: \blue{19.082} \teal{\plusminus} \blue{0.1}
      - \violet{Luminous}:
        x: \blue{(-1.4772, 0.5532)} \teal{\plusminus} \blue{0.002}
        mag: \blue{19.149} \teal{\plusminus} \blue{0.1}
      - \violet{Luminous}:
        x: \blue{(-0.9377, -1.6147)} \teal{\plusminus} \blue{0.002}
        mag: \blue{19.196} \teal{\plusminus} \blue{0.1}
\end{Verbatim}

\end{appendix}

\end{document}